\documentclass[12pt]{iopart}
\usepackage{iopams}  
\usepackage{graphicx}
\begin{document}

\title[Local structure correlations in cyclohexane]{Local structure correlations in plastic cyclohexane -- a Reverse Monte Carlo study}

\author{Nicholas P Funnell$^{1,2,\ast},$ Martin T Dove$^3$, Andrew L Goodwin$^1$, Simon Parsons$^2$ and Matthew G Tucker$^4$}
\address{$^1$ Department of Chemistry, University of Oxford, Inorganic Chemistry Laboratory, South Parks Road, Oxford, OX1 3QR}
\address{$^2$ EaStChem and Centre for Science at Extreme Conditions, University of Edinburgh, West Mains Road, Edinburgh, EH9 3JJ}
\address{$^3$ School of Physics and Astronomy, Queen Mary University of London, Mile End Road, London, E1 4NS}
\address{$^4$ ISIS Facility, Rutherford Appleton Laboratory, Harwell Science and Innovation Campus, Didcot, Oxfordshire, OX11 0QX}

\ead{nicholas.funnell@chem.ox.ac.uk}

\begin{abstract}
Two solid phases of cyclohexane have been investigated over a temperature range spanning 13 to 266\,K on a powdered, perdeuterated sample using neutron total scattering. Phase II has an ordered structure (\textit{C}2/\textit{c}) that forms below 186\,K. Between 186 and 280\,K it exists as a plastic solid -- phase I (\textit{Fm}3\textit{m}), where the molecules are rotationally disordered about the lattice points of the face-centred cubic cell.  Data-dependent atomistic configurations that represent the `instantaneous' crystal structure have been generated from the total scattering data using Reverse Monte Carlo refinement. Analysis of local structure reveals that instantaneous distortions in phase I resemble the average structure of phase II.
\end{abstract}


\maketitle

\section{Introduction}
Understanding the behaviour of matter in the solid state is crucial for the rational development of compounds such as functional materials or pharmaceuticals where intentional induction or prevention of a phase transition is key to the performance of the compound \cite{Sun_2008,Caravati_2010,Chemburkar_2000}. One approach to understanding this behaviour is through quantification of the relative stabilities of polymorphic crystals, or two structures either side of a phase transition, in terms of interaction energies and/or thermodynamic parameters \cite{Wood_2006,Rivera_2008,Funnell_2011,Johnstone_2008}. However by taking this approach, one aspect of phase behaviour is overlooked -- how the molecules interact locally. Average-structure models, such as those obtained from a typical crystallographic refinement are useful for observing how a material behaves over the duration of data collection, but they do not reveal how the molecules interact at a local level, nor how they influence neighbouring molecules as the substance approaches a phase transition.

Neutron total scattering presents an appealing approach for monitoring phase transitions given its capability of simultaneously measuring both the average (crystal) structure and the instantaneous local environment around each atom. With respect to the solid state, the total scattering field has been mostly limited to extended inorganic structures, such as strontium titanate \cite{Hui_2005}, quartz \cite{Tucker_2000}, cristobalite \cite{Tucker_2001} and the Pb$_{2}$Ru$_{2}$O$_{6.5}$ oxide pyrochlore \cite{Shoemaker_2011}; there are comparatively few experiments that focus on molecular systems and even fewer that aim to model the structure from the scattering data. Molecular systems arguably present a greater challenge, given that intermolecular interactions are much weaker than covalent bonds and so each molecule and, by extension, each atom in a molecule is inherently more predisposed towards occupying a wider range of positions. Whilst relatively simple molecular systems such as SF$_6$ \cite{Dove_2002} and CBr$_4$ \cite{Temleitner_2010} have been studied, where orientational disorder in each of these was reproduced in atomistic models through direct refinement against the scattering data, molecules of increasing complexity are starting to be explored. Qualitative and quantitative comparisons of the measured pair distribution functions from amorphous and crystalline forms of carbamazepine and indomethacin have been performed, with the intention of developing a method for `fingerprinting' amorphous and nanocrystalline pharmaceuticals \cite{Dykhne_2011,Billinge_2010}. One of the challenges posed by large molecules is the difficulty in distinguishing intra from intermolecular contacts between atoms of the same type when they are separated by similar distances. A study of the $\rho$-terphenyl molecule aimed to tackle this problem by identifying differences in pair distribution function peak shapes to differentiate between intra and intermolecular correlations. This information was used to make adjustments to an atomistic model so as to be consistent with the experimental data \cite{Rademacher_2012}. To the best of our knowledge, this study of $\rho$-terphenyl is the first `sizeable' organic molecule to be studied using a combined total scattering/atomistic modelling approach.

Cyclohexane, the focus of this present investigation, is an important molecular system that has also been studied using neutron total scattering \cite{Farman_1987,Farman_1991,Farman_1996}. Of fundamental interest is its thermally-induced solid-solid phase transition at 186\,K. Below this temperature it exists as an ordered crystalline phase (\textit{C}2/\textit{c}, hereafter referred to as `phase II'); whereas between 186 and 280\,K it is a plastic solid, in which the molecules are rotationally disordered about the lattice points of a FCC cell (\textit{Fm}3\textit{m}, `phase I'). The structures of these phases are shown in Figure 1. The packing arrangement in both phases has been previously identified through single-crystal X-ray diffraction, where single crystals of the plastic phase were obtained by growth near the transition point, using a zone melting technique \cite{Kahn_1973}. The same study showed that phase I can also been obtained as a Ôplastic crystal glassÕ at 77\,K when quenched. Raman scattering, molecular dynamics (MD) simulations and NMR experiments have all been performed with the aim of identifying the precise nature of the disorder both in phase I and in the liquid above 280\,K \cite{Bansal_1979,Brodka_1992,Ehrenberg_1992,McGrath_1997}. This was also the intention of a series of neutron total scattering studies which predominantly focussed on the plastic and liquid phases \cite{Farman_1987,Farman_1991,Farman_1996}. However, owing to instrument limitations at the time, data were only collected to a $Q_\mathrm{max}$  = 16\,{\AA}$^{-1}$. Modern total scattering instruments are now capable of collecting data to much higher resolution, \textit{ca}.~50\,{\AA}$^{-1}$, which mitigates the effects of Fourier truncation and consequently makes it easier to discern real peaks from Fourier ripples in a real-space distribution of interatomic distances. By collecting data to higher resolution, we are able to use the data to drive refinement of 3-D atomistic models from which orientational and spatial correlation functions can be extracted. The focus on the transformation between phases I and II rather than the melting transition aims to explore the relationship between ordered and disordered phases. 

\begin{figure}
\begin{indented}
\item \includegraphics[width=10cm]{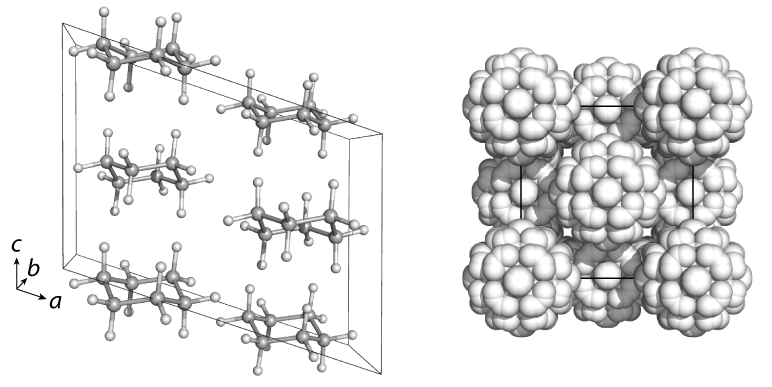}
\end{indented}
\caption{\label{Figure_1.png} The structure of cyclohexane in phase II (left) and phase I (right). Only deuterium atoms are visible in phase I.}
\end{figure}

The total scattering data reported in this manuscript have been modelled using a Reverse Monte Carlo (RMC) \cite{McGreevy_1988} approach, producing atomistic structures that are consistent with both the instantaneous local environment and the average structure, as would be seen by traditional Rietveld refinement. Farman \textit{et al} suggest that RMC refinement could potentially be used to understand the phase behaviour of cyclohexane further, although they point out that chemically unrealistic atomic configurations are frequently generated \cite{Farman_1996}. Following advances in RMC software over the last couple of decades, this is now readily prevented with judicious use of geometry-based restraints.

We report our analysis of high-resolution neutron total scattering data ($Q_\mathrm{max}$ = 45\,{\AA}$^{-1}$) for cyclohexane-$d_{12}$, over a temperature range spanning phases I and II. From the resulting RMC-refined models, we are able to extract correlation functions which reveal translational and rotational relationships between neighbouring pairs of molecules in both phases.

\section{Experimental}
\subsection{Sample preparation}
Cyclohexane-$d_{12}$ (99.5\% D), obtained from CDN isotopes, was frozen with liquid nitrogen and cold-ground into a homogenous powder at 77\,K, in a nitrogen atmosphere. Details of the cold-grinding apparatus are given elsewhere \cite{Ibberson_1996}.

\subsection{Neutron total scattering}
The cyclohexane sample was packed into a thin-walled vanadium can and mounted on a CCR device to control the sample temperature. Total scattering data were collected on the General Materials Diffractometer (GEM) at the ISIS pulsed neutron and muon source \cite{Williams_1998} using the time-of-flight (ToF) method at the following temperatures: 13, 75, 126, 176, 206 and 266\,K. The measured differential cross-section data were processed using GUDRUN \cite{McLain_GUDRUN}, to correct for background scattering, multiple scattering, Placzek inelasticity and beam attenuation by the sample container, giving the normalised total scattering functions $F(Q)$ ($Q_{\mathrm{max}}$ = 45\,{\AA}$^{-1}$) and the corresponding pair distribution functions $D(r)$. For reference, a thorough comparison of commonly used total scattering formalisms is given by Keen \cite{Keen_2001}.

The Bragg contribution to the total scattering patterns was analysed separately using conventional average structure refinement as implemented in GSAS \cite{Larson_GSAS,Toby_2001}. Rietveld refinement of phase II used the atomic coordinates of the cyclohexane structure deposited in the Cambridge Structural Database with the Refcode `CYCHEX' as an initial starting model \cite{Kahn_1973,Bruno_2002}. Refinement was carried out against data collected on all detector banks with exception of the 2$\theta$ = 5--12\,$^\circ$ bank which contained very few Bragg peaks over the accessible range of \textit{d}-spacing and so was excluded. The fit and data corresponding to the 2$\theta$ = 50--74\,$^\circ$ bank were used to generate the files required for the Bragg profile component in RMC refinement. Rietveld refinement of phase I utilised a Frenkel model, where the cyclohexane molecule was modelled over 24 orientations, effectively rotating about its centre of mass. Kahn \textit{et al} found this model gave a marginally better fit to X-ray single crystal data than a completely isotropically disordered model \cite{Kahn_1973}. The refinement was carried out against data collected on the 2$\theta$ = 50--74\,$^\circ$ detector bank only, which was then used to generate the files for RMC refinement.  Rietveld fits to the data at 13, 176 and 266\,K are shown in Figure 2.

\begin{figure}
\begin{indented}
\item \includegraphics[width=8.0cm]{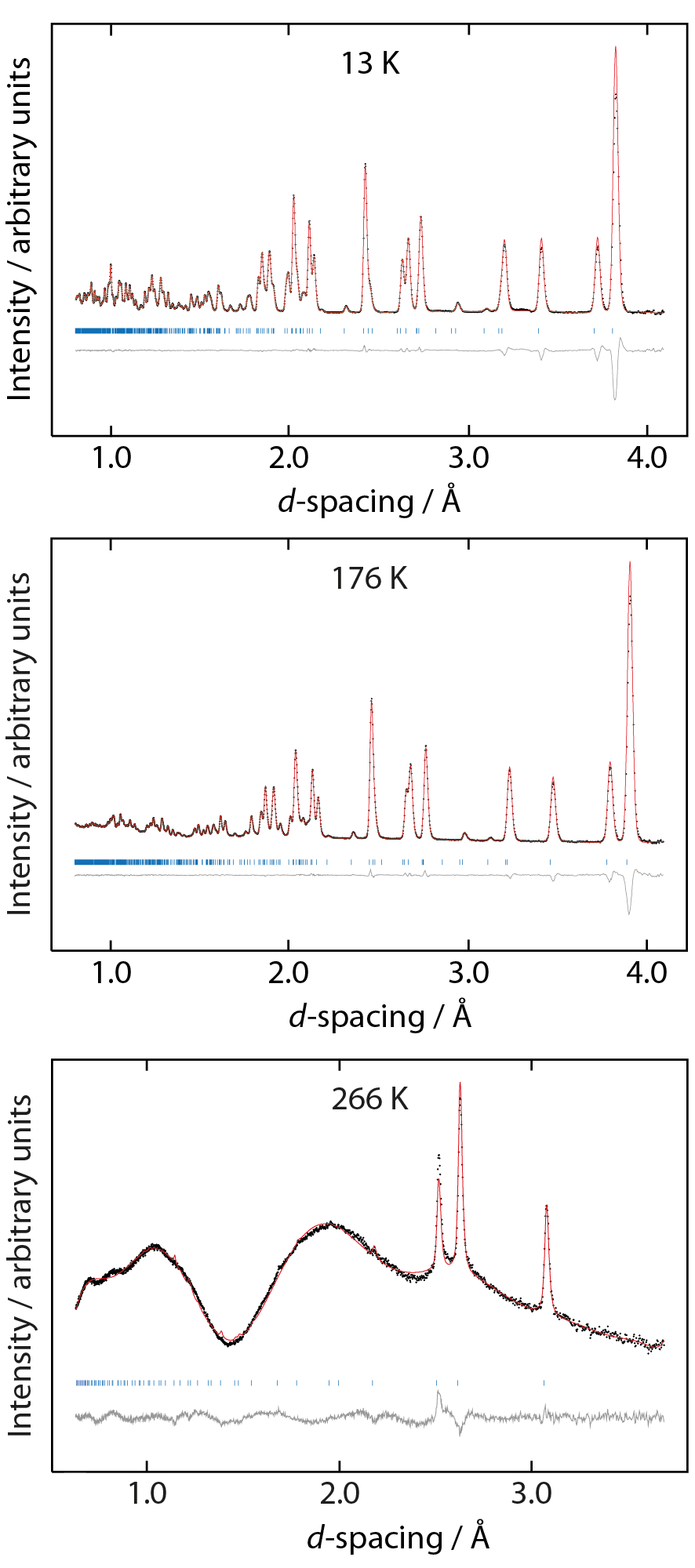}
\end{indented}
\caption{\label{Figure_2.png} Rietveld fits to the phase II data at 13 and 176\,K and phase I at 266\,K}
\end{figure}

\subsection{Starting models for Reverse Monte Carlo refinement}
Starting models for the RMC refinement at each temperature point were obtained via MD simulations using the DISCOVER software package implemented in Accelrys Materials Studio. The cvff forcefield was used with an NPT ensemble \cite{Dauber_1988}. Simulations were carried out for 20\,ps at intervals of 0.5\,fs using a Berendsen thermostat and a Parrinello barostat \cite{Berendsen_1984,Parrinello_1982}.

A $6\times6\times6$ supercell was created for each temperature point, consisting of 15552 atoms. The phase II models were obtained by using the Rietveld-refined 13\,K crystal structure as a starting set of coordinates and then performing MD simulations on these at each experimentally measured temperature. The dimensions of the resulting supercells were then manually adjusted to those of the relevant, experimentally observed, crystal structure supercell before undergoing RMC refinement. Construction of the starting models for phase I was a little more elaborate: molecules were manually placed on the FCC lattice points of a cell with \textit{P}1 symmetry and dimensions corresponding to the plastic crystal at 206\,K. The molecules were arbitrarily orientated in positions that appeared visually to be sterically favourable. The geometry was coarsely optimised using DMol$^3$ \cite{Delley_1990} (final $\Delta E = 4 \times 10^{-5}$ Ha) and then MD simulations on a $6\times6\times6$ supercell were performed. The purpose of the geometry optimisation was to relieve instabilities in the initial cycles of the MD simulation, which were observed if the geometry was not optimised. The resulting supercell was then manually adjusted to have the experimentally observed cell dimensions prior to RMC refinement.

\subsection{Reverse Monte Carlo refinement}
The program RMCProfile \cite{Tucker_2007} was used to simultaneously fit calculated $F(Q)$, $D(r)$ and Bragg peak intensities to the experimentally observed patterns by randomly moving atoms in the supercell. Moves were accepted based on minimisation of the function

\ 
\begin{equation}
\chi^2_{\mathrm{RMC}} = \sum_m \chi^2_\mathrm{m}
\end{equation}
\ 

\noindent where $\chi_\mathrm{m}^2$ corresponds to each data set being refined. The individual $\chi^2$ functions being minimised were:

\ 
\begin{equation}
\chi^2_{QF(Q)} = \sum_j [QF_{\mathrm{calc}}(Q_j) - QF_{\mathrm{exp}}(Q_j)]^2\sigma ^{-2}_{QF(Q)}
\end{equation}
\begin{equation}
\chi^2_{D(r)} = \sum_j [D_{\mathrm{calc}}(r_j)-D_{\mathrm{exp}}(r_j)]^2\sigma ^{-2}_{D(r)}
\end{equation}
\begin{equation}
\chi^2_{\mathrm{profile}} = \sum_j [I^{\mathrm{calc}}_{\mathrm{profile}}(t_j) - I^{\mathrm{exp}}_{\mathrm{profile}}(t_j)]^2\sigma ^{-2}_\mathrm{profile}
\end{equation}
\begin{equation}
\chi^2_{\mathrm{BS}} = \frac{1}{k_\mathrm{B}T} \sum_l D_l\{1-\exp[-\alpha(r-r_0)]\}^2
\end{equation}
\begin{equation}
\chi^2_{\mathrm{BB}} = \frac{1}{k_\mathrm{B}T} \sum_l K_l(\cos\theta - \cos\theta_0)^2
\end{equation}
\ 

\noindent where $\sigma$ is a weighting parameter, defined for each separate set of data and $I_\mathrm{profile} (t)$ corresponds to the Bragg peaks, fitted as a function of neutron flight time. The $\chi_\mathrm{BS}^2$ and $\chi_\mathrm{BB}^2$ functions are used for geometric restraints (discussed further below), where $D$ is the energy required to break a bond, $\alpha$ specifies the curvature of the potential energy function, $r_0$ is the specified bond length, $\theta_0$ is the specified bond angle and $K$ is the bond angle energy term. All moves that decreased the value of $\chi^2$ were automatically accepted and those that increased $\chi^2$ were accepted within the probability limit

\ 
\begin{equation}
P=\exp(-\Delta\chi^2/2)
\end{equation}
\ 

\noindent to avoid becoming trapped in local minima. Refinements of the datasets were carried out until no further improvements in the value of $\chi^2$ were seen, requiring \textit{ca}.~100 accepted moves per atom.

In order to generate chemically reasonable atomic configurations, forcefield-based distance and angle restraints were imposed on the molecular geometry as otherwise RMCProfile produced erroneous atomic connectivity. Given the rigidity of the cyclohexane molecule, the intramolecular bond distances and angles should not change significantly over the temperature range investigated here and so application of these restraints was quite reasonable. The values used for the bond distances ($r_0$ in (5)) and angles ($\theta_0$ in (6)) were set to those observed in the crystal structure at 13\,K. $D_l, \alpha$ and $K_l$ were set to the default values in the MM3 database. Closest approach restraints were also employed, preventing D--D, C--D and C--C atom pairs from moving within 1.4, 0.9 and 1.3\,{\AA} of each other respectively. These distances were intentionally set below the realistic closest approach distances seen experimentally (1.7, 1.1 and 1.5\,{\AA}) to try and reduce the level of structural bias introduced into the model as much as possible.

For each phase II dataset, RMCProfile used the scale factor determined in the corresponding GSAS refinement to fit the Bragg peak intensities, and this was held fixed for the duration of the run. RMC refinement of the phase I data was more problematic owing to the difficulty in obtaining a reliable scale factor from the Rietveld refinement, as it was strongly correlated with the atomic isotropic thermal parameters. Thus the overall scale factor for the Bragg intensities was also optimised during the RMC refinement. 

\subsection{Other programs used}
Atomic configurations were visualised using CrystalMaker{\textsuperscript \textregistered} 8.6 \cite{Crystalmaker} and Mercury 3.0 \cite{Macrae_2008}. Structural figures were created using Atomeye \cite{Li_2003} and Pymol \cite{Pymol}. 3-D correlation function data were visualised using Paraview \cite{Henderson_2007}. Cambridge Structural Database searches were carried out using CONQUEST with database updates up to May 2012 \cite{Bruno_2002}.

\section{Results}
\subsection{Initial agreement with data}
Prior to RMC refinement, $D(r)$, $F(Q)$ and Bragg profile patterns were calculated from the starting models and compared with the experimental data. The calculated and experimental $D(r)$ data at 13, 176 and 266\,K are shown in Figure 3. Across all the temperatures, the MD simulations result in structures that are too strongly correlated at \textit{ca.}~$r <$ 5\,{\AA}, and at 13 K, this excessive correlation extends over the entire range of real space being explored. This is likely a consequence of the MD simulations preserving `ideal' bond lengths --- an indication that the bond stretching and bending force constants are too high. However an additional contribution to the peak width in the experimental data at low \textit{r} also arises from the treatment of the total scattering data itself, as Fourier transformation of the $F(Q)$ data over a finite $Q$ range introduces $r$-dependent peak broadening. 

At 176\,K there is markedly better agreement between the MD model and experimental data above 5\,{\AA}, but at 266\,K the agreement with respect to peak positions is rather poor, particularly between 2--5\,{\AA} --- the region corresponding to nearest neighbour interactions.  

\begin{figure}
\begin{indented}
\item \includegraphics[width=\textwidth]{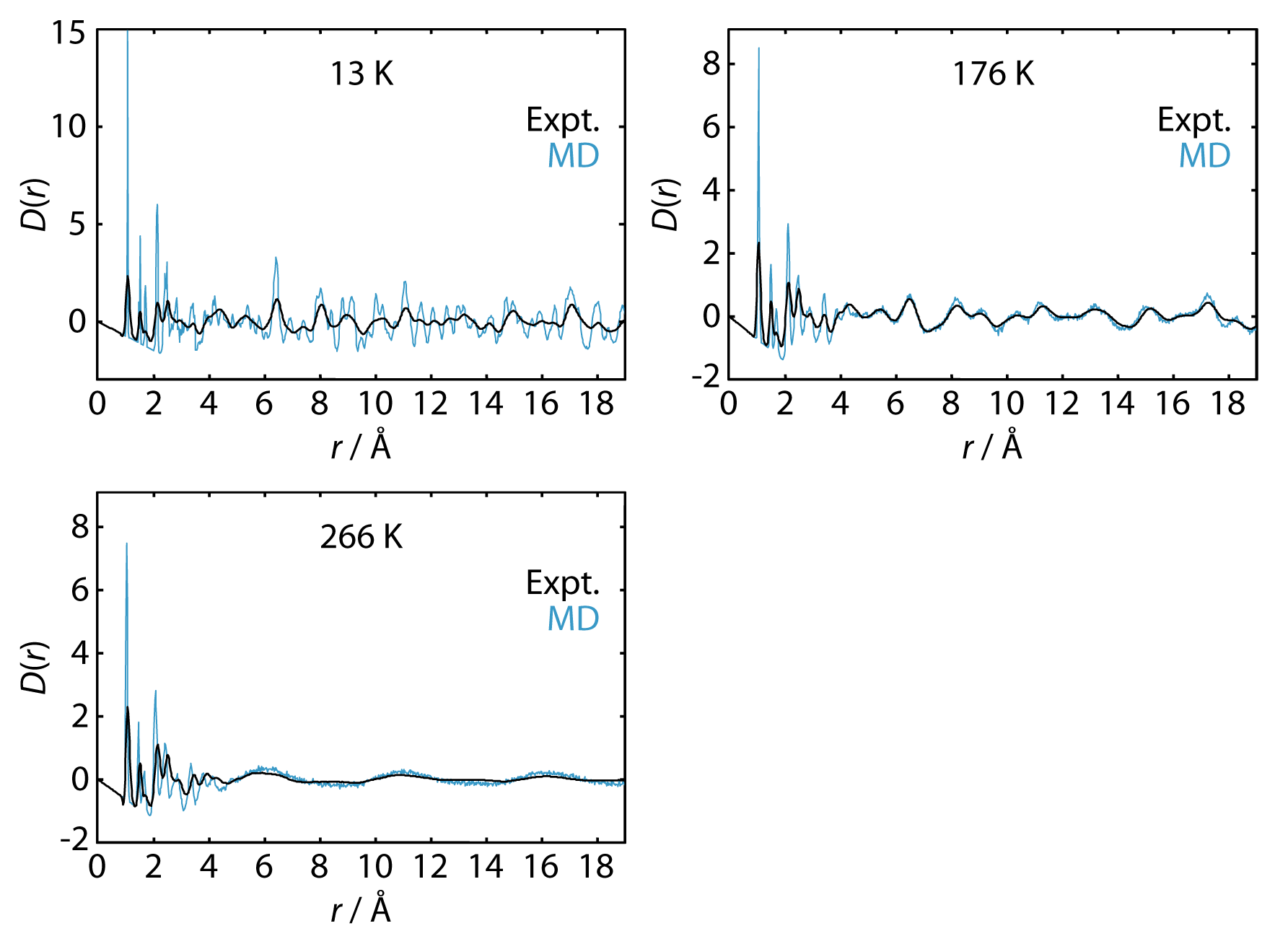}
\end{indented}
\caption{\label{Figure_3.png} Initial level of agreement between the experimental $D(r)$ data and the simulated MD models at 13, 176 and 266 \,K.}
\end{figure}

\subsection{RMCProfile refinement} 
The final RMCProfile fits to the $D(r)$, $F(Q)$ and Bragg profile data are shown in Figure 4 for the data at 13, 176 and 266\,K. Plots of the final fits to the data at all temperature points investigated can be found in the Supplementary Information. 

\begin{figure}
\begin{indented}
\item \includegraphics[width=\textwidth]{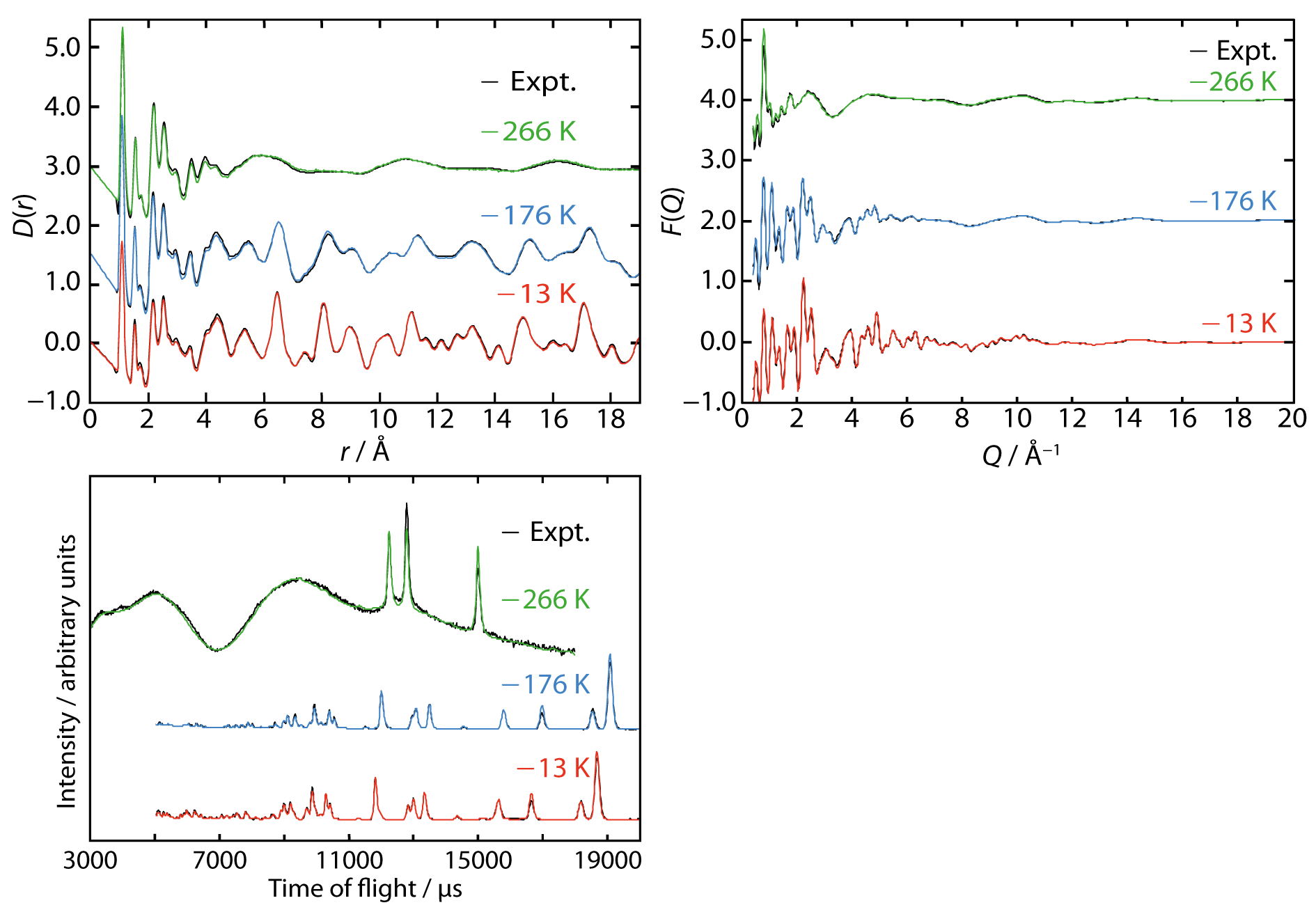}
\end{indented}
\caption{\label{Figure_4.png} RMC fits to the $D(r)$ (upper left), $F(Q)$ (upper right) and Bragg profile (lower left) data at 13, 176 and 266\,K. The $F(Q)$ data are shown to $Q_\mathrm{max}$ = 20\,{\AA}$^{-1}$ to show the low-$Q$ region more clearly. Plots showing the full $Q$ range can be found in the S.I. The Bragg profile data at 266\,K have been scaled up by a factor of 20 to be visually comparable with the other data}
\end{figure}

Figure 4 shows that the fits to all phase II data are generally very good. The $D(r)$ plots at $r \leq$ 5\,{\AA} change very little as a function of temperature since the major contribution to this region of the pattern is from intramolecular contacts, which are not expected to vary significantly. The intermolecular contacts --- the shortest of these being between deuterium atoms --- only start to contribute to the $D(r)$ above \textit{ca}.~1.8\,{\AA}. A very small degree of peak broadening is present in the low-$r$ region, reflecting the effect of thermal energy on the magnitude of bond stretching. Conversely, the region above \textit{ca}.~5\,{\AA} exhibits significant peak broadening when temperature is raised, representing increased deviation of the instantaneous molecular positions from those of the average positions. This can be solely ascribed to intermolecular contacts given that the longest intramolecular distance is \textit{ca}.~5\,{\AA} between the equatorial deuterium atoms on opposite sides of the cyclohexane ring.

We make a brief comment here regarding the use of the $D(r)$ normalisation in RMC refinement. Preliminary refinements with the same data and constraints, not reported here, were performed with the $G(r)$ normalisation instead. It was found that although satisfactory fits were obtained for all data, when these fits were recalculated with $D(r)$ it was apparent that the phase I atomic configurations did not adequately reflect the experimental $D(r)$ at large $r$ ($>$ 10\,{\AA}). This had the effect that phase I models with some remnant artificial orientational periodicity could still yield good fits to the data. 

Returning to the fits in Figure 4, the difference between phase I and phase II most obviously manifests itself in the Bragg profile data -- there are only three significant Bragg peaks sat on a large background of modulated diffuse scattering, which is consistent with the onset of rotational disorder in the structure. This point is also clearly evident in the very diffuse peaks in the $D(r)$ above 5\,{\AA} and the disappearance of peaks in the $F(Q)$. Below 5\,{\AA}, the $D(r)$ data are similar to phase II as the connectivity of the molecule remains unchanged. 

The positions of the broad features in the $D(r)$, corresponding to intermolecular correlations, in the 266\,K data are comparable with those reported by Farman at 263\,K (5.9, 10.9, 16.2 and 21.6\,{\AA} in our data \textit{c.f}. 6.0, 11.1, 16.3 and 21.5\,{\AA} in Farman's data) \cite{Farman_1991}. It was noted by Farman that the first purely intermolecular peak at 6.0\,{\AA} is at lower $r$ than might be expected as the 8.6836\,{\AA} FCC cell edge means that the shortest contact between molecular centroids should be \textit{ca}.~6.14\,{\AA}. In our data, the apex of this very diffuse peak, which ranges between 4.72 and 7.50\,{\AA} actually appears at an even shorter distance of 5.92\,{\AA}. This all points toward an increased need for neighbouring molecules to interact in a correlated manner whereby molecular orientations are coupled to nearest-neighbour separation distances.

Aside from peaks in the $D(r)$ that can be attributed to intramolecular contacts, there are additional peaks at longer $r$ ($>$ 5\,{\AA}) common to both phases that only differ in position by a maximum of \textit{ca}.~0.4\,{\AA}. It is known from the earlier neutron total scattering experiments \cite{Farman_1996} that the structure of the plastic phase is dependent on the thermal history of the sample and so this suggests that even though the phase I structure may appear rotationally disordered, there are strong local correlations present similar to those seen in phase II.

\subsection{Distribution of atomic positions}
In order to gain a clear picture of the atomic site distribution across all temperatures, 10 RMC refinements were carried out at each temperature point, and the resulting configurations combined. This allows improved estimation of uncertainties in derived parameters such as correlation functions. Figure 5 shows the distribution of carbon atom positions in the unit cell as a function of temperature, compared with the thermal ellipsoids of the average crystal structure plotted at the 99\% probability level. In the interest of improved statistical significance all subsequent analysis reported here has been carried out on this ensemble of 10 RMC configurations for each temperature.

\begin{figure}
\includegraphics[width=\textwidth]{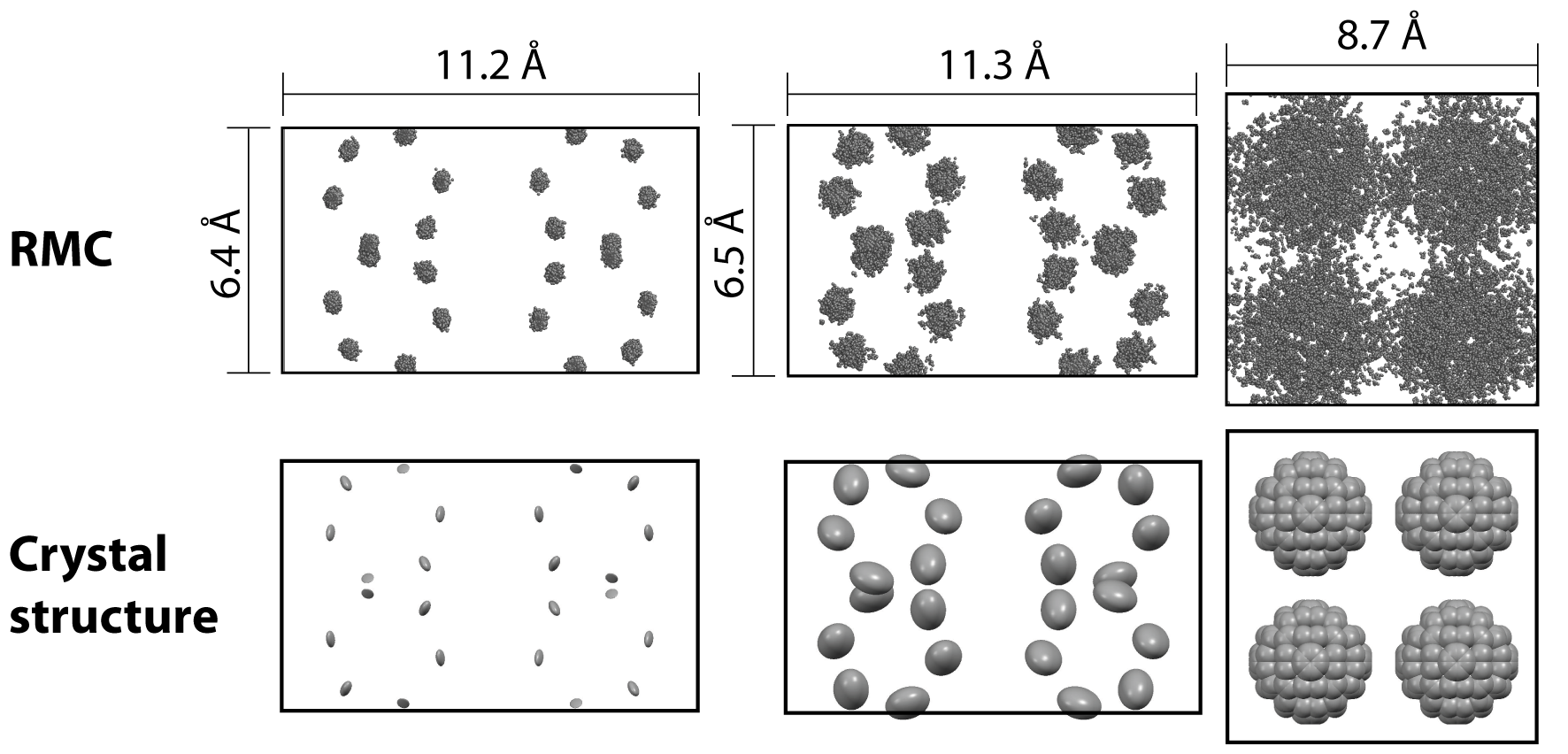}
\caption{\label{Figure_5.png} Comparison of the instantaneous distribution of carbon atoms with the average crystal structures as viewed along the \textit{c}-axis at 13 (left), 176 (centre) and 266\,K (right). Thermal ellipsoids in the lower row are displayed at the 99\% level. Deuterium atoms are omitted for clarity. The two phases are not shown to scale with respect to each other and so cell constants have been included. The origin of the unit cell in the 266\,K crystal structure (lower right) has been shifted to aid visual comparison with the corresponding instantaneous structure.}
\end{figure}

In phase II, the distribution of carbon atoms closely resembles the anisotropic displacement parameters in the Rietveld-refined crystal structure, reproducing the effects of increasing temperature on atomic position. In phase I, the distribution of atoms about each lattice point is approximately spherical although the fit to the Bragg profile data is not as good as that of phase II --- the intensity of one peak is under-fitted and another over-fitted.  We ascribe this to the probable challenge presented to the RMC process by the model itself. The combination of close molecular packing and minimum distance constraints may make it difficult to rotate a molecule, as it necessarily requires other neighbouring molecules to rotate in a cooperative manner. Thus, the final molecular orientations are more likely the result of the MD simulations instead. Evidence for this is presented in the Discussion. However, the RMC process greatly improves upon the separation distances between molecules, as shown by the initially poor agreement between the calculated and experimental $D(r)$'s (Figure 3) and the excellent agreement obtained when fitted (Figure 4). Whilst the refined distribution of carbon atoms mostly conforms to the FCC symmetry restrictions, there are numerous carbon atoms that are not located within one of the four `main sites' of atomic distribution, shown in Figure 5. These correspond to a few molecules in each of the datasets that have moved away from the FCC sites during the RMC refinement. These spatial deviations may not be structurally unrealistic given that the position of the first intermolecular peak in the $D(r)$ appears at a shorter distance than the smallest intermolecular separation in the average structure.

\section{Discussion}
\subsection{Correlated motion}
In phase I each molecule should not be able to tumble independently of its nearest neighbours, simply because there is insufficient space to do so \cite{Farman_1991}. In order for a molecule to tumble, the closest molecules must spatially accommodate it, but this necessarily has implications for successive shells of nearest neighbours. An advantage of fitting an atomistic model to the data presents itself here --- namely features such as the spatial distribution of molecules and their orientations can be extracted. 

An initial assessment of intermolecular D{\dots}D contacts shows that the vast majority of molecules do not move unfavourably close to one another, avoiding the occurrence of contacts below 1.7\,{\AA}; the closest approach distance for intermolecular hydrogen atoms identified by Wood \textit{et al} \cite{Wood_2008} for small organic molecules up to pressures of 10\,GPa. A small number of molecules approach each other at distances as low as 1.4\,{\AA} --- the lowest limit permitted by the RMC refinement. Figure 6 shows a radial distribution function of the contacts as a function of temperature, clearly showing that these unrealistic super-short distances are the exception rather than the rule, where the `real' minimum distance actually appears to be closer to \textit{ca.}~1.9\,{\AA} instead. Since the majority of molecules obey the 1.7\,{\AA} limit, this immediately shows pairwise correlated motion exists to some extent.

In order to obtain a quantitative understanding of the orientational correlations between pairs of molecules, we proceeded to analyse the RMC configurations purely in terms of the orientation of each cyclohexane ring and the location of its centre of mass. For convenience, we represent the orientations as vectors \textbf{S} oriented normal to the mean plane of the cyclohexane ring, noting that all subsequent calculations are invariant with respect to inversion. Figure 7 shows representative sections of the atomic configurations at 13, 176 and 266\,K with the vectors represented by double-ended red arrows to reflect this inversion invariance. Orientational correlations between cyclohexane molecules can then be quantified by the correlation function $\chi = 2|\textbf{S}{\cdot}\textbf{S}^{\prime}| -1$. Here $\textbf{S}$ and $\textbf{S}^{\prime}$ represent the individual molecular orientations for any given pair of neighbouring molecules. The value of $\chi$ is equal to 1 where the molecules are parallel and $\chi$ = $-$1 when they are perpendicular to one another. Given the likelihood that the molecular orientations are strongly coupled to separation distance, the relative spatial positions of each molecular pair were also considered through definition of a centroid--centroid vector, \textbf{r}, forming the dot products $|\textbf{S}{\cdot}\textbf{r}|$ and $|\textbf{S}^{\prime}{\cdot}\textbf{r}|$. Figure 8 shows the four extreme orientational and translational relationships that can exist between two of the molecules and the corresponding values of $\chi$, $|\textbf{S}{\cdot}\textbf{r}|$ and $|\textbf{S}^{\prime}{\cdot}\textbf{r}|$ and $|\textbf{r}|$ in each scenario.

\begin{figure}
\begin{indented}
\item \includegraphics[width=10cm]{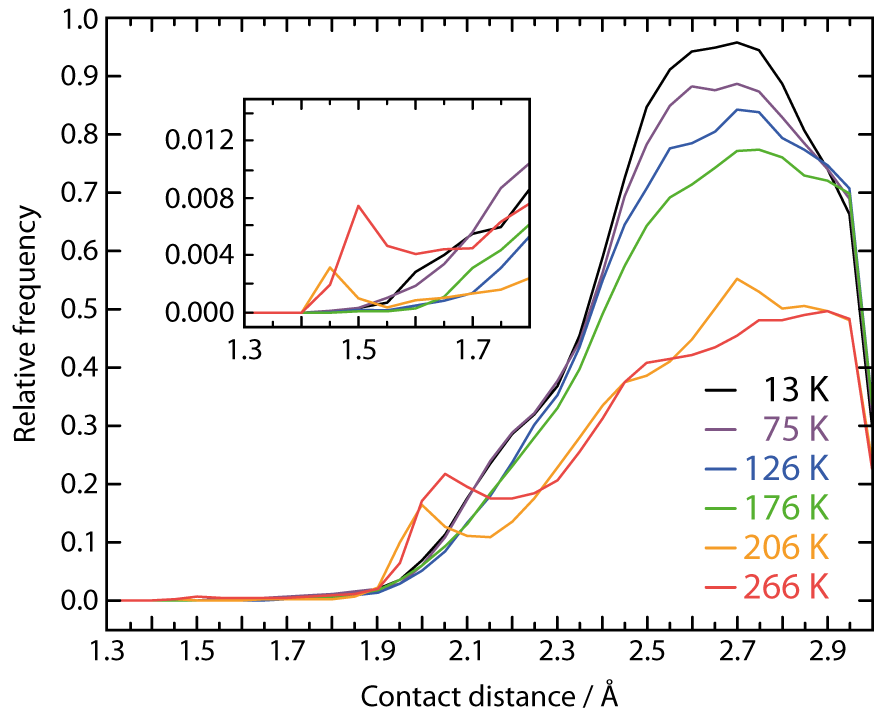}
\end{indented}
\caption[width=\textwidth]{\label{Figure_6.png} Partial radial distribution function for intermolecular D$\dots$D contacts, colour coded by temperature. The tick marks on the x-axis indicate the centre of bins with widths of 0.5\,{\AA}. The inset plot shows a magnified region of the histogram between 1.3 and 1.8\,{\AA}.} 
\end{figure}

\begin{figure}
\includegraphics[width=\textwidth]{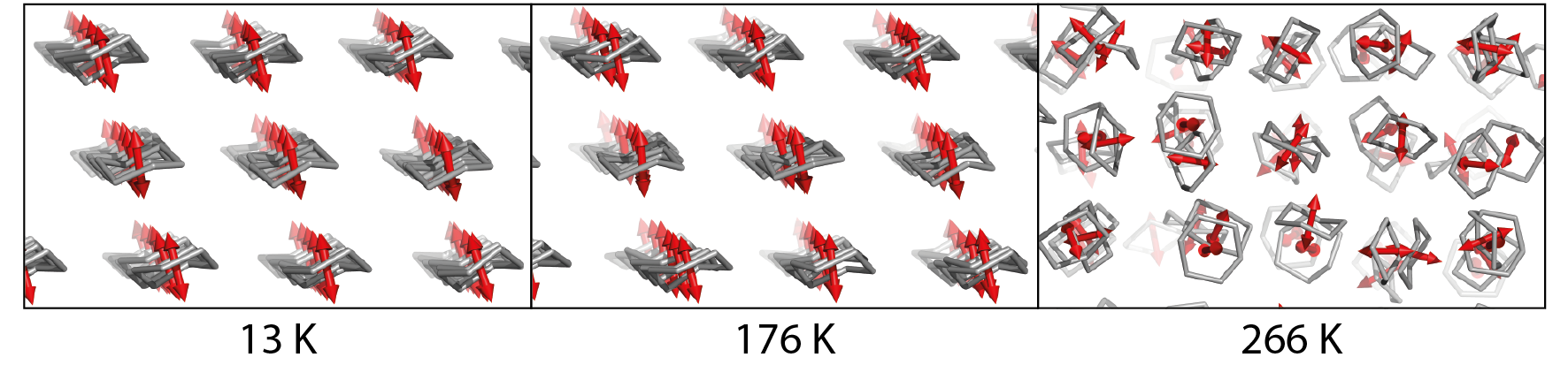}
\caption[width=\textwidth]{\label{Figure_7.png} Representative sections of the RMC atomic configurations at 13, 176 and 266\,K. The arrows represent the axis perpendicular to the mean plane of the molecules. Deuterium atoms are omitted for clarity.}
\end{figure}

\begin{figure}
\begin{indented}
\item \includegraphics[width=8cm]{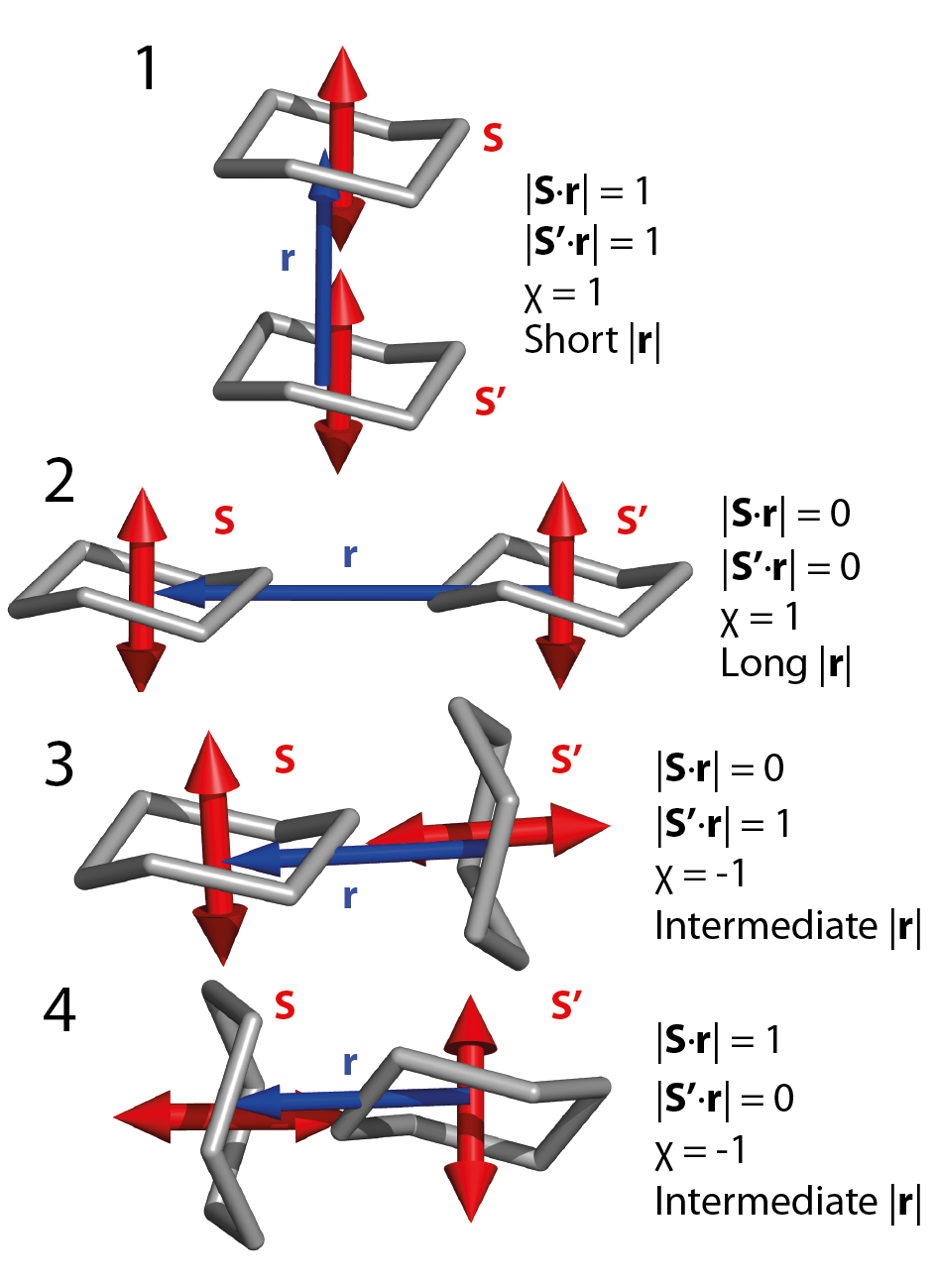}
\end{indented}
\caption[width=8cm]{\label{Figure_8.png} The four molecular pair arrangements that result in the most extreme values of $\chi$, $|\textbf{\textrm{S}}{\cdot}\textbf{\textrm{r}}|$ and $|\textbf{\textrm{S}}^\prime{\cdot}\textbf{\textrm{r}}|$. Scenarios 3 and 4 are equivalent to each other.}
\end{figure}

To investigate the extent of correlation between molecular orientations and their separation distance, the dependence of $\chi$ on $|\textbf{S}{\cdot}\textbf{r}|$, $|\textbf{S}^{\prime}{\cdot}\textbf{r}|$ and $|\textbf{r}|$ was calculated at 13, 176 and 266\,K, and represented using isosurfaces, plotted in Figure 9. The red and blue surfaces in the Figure correspond to positive and negative values of $\chi$, respectively; i.e. to molecular pairs that tend towards parallel (red) or perpendicular (blue) orientations. It is important to emphasise that the surface colours do not mean that the molecules adopt \textit{only} the orientations shown in Figure 8 but rather any orientation between 45 and 0$^{\circ}$, or 45 and 90$^{\circ}$ with respect to each other. The value of $\chi$ is calculated in both directions along the molecular separation vector and so the surfaces are inherently symmetrical about the diagonal between the $|\textbf{S}{\cdot}\textbf{r}|$ and $|\textbf{S}^{\prime}{\cdot}\textbf{r}|$ axes.

\begin{figure}
\includegraphics[width=\textwidth]{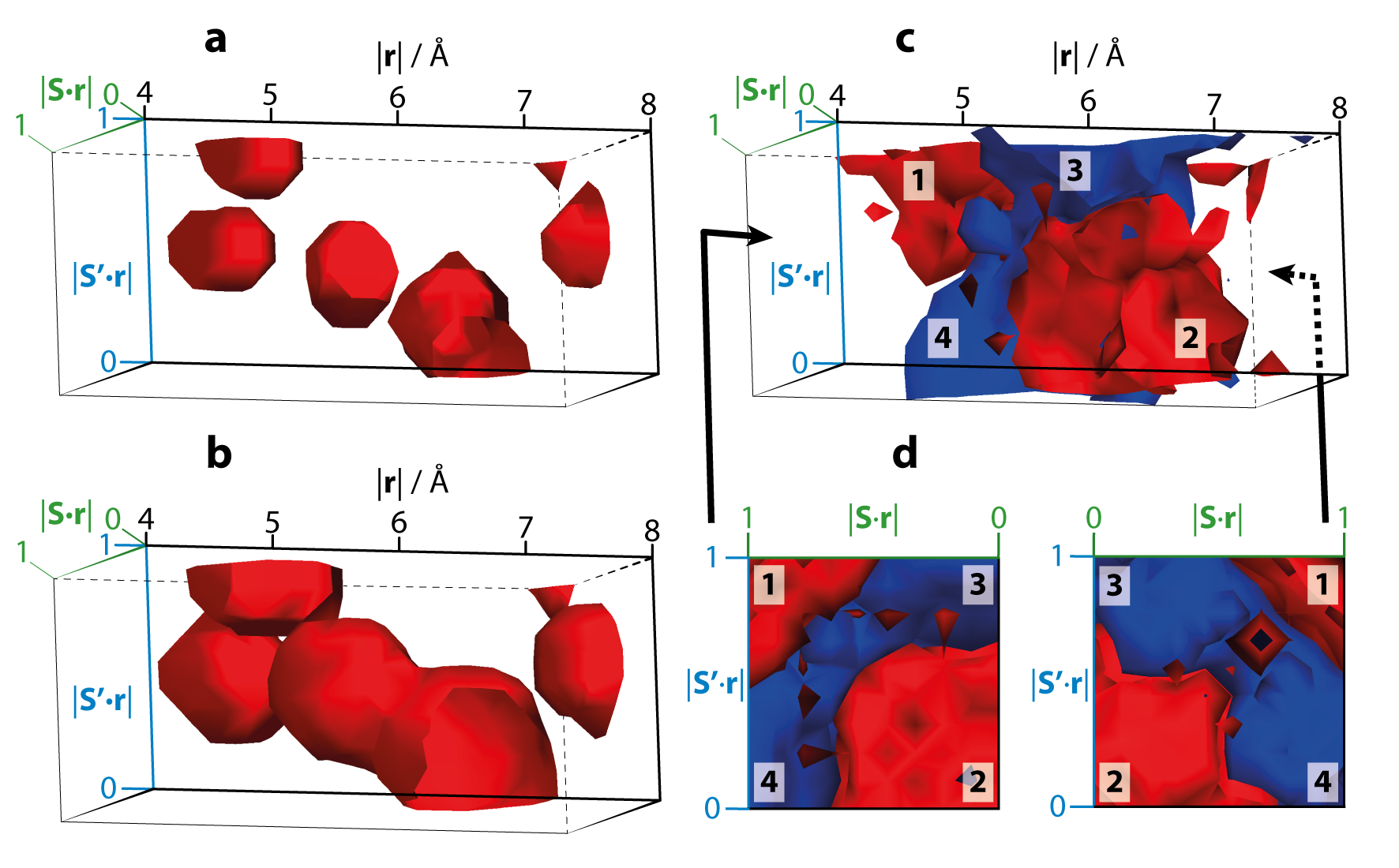}
\caption[width=\textwidth]{\label{Figure_9.png} Isosurfaces showing the correlation function, $\chi$ for (\textbf{a}) 13\,K, (\textbf{b}) 176\,K and (\textbf{c}) 266\,K as a function of $|\textbf{\textrm{r}}|$, $|\textbf{\textrm{S}}{\cdot}\textbf{\textrm{r}}|$ and $|\textbf{\textrm{S}}^\prime{\cdot}\textbf{\textrm{r}}|$ along the horizontal, diagonal and vertical axes, respectively. \textbf{d} shows \textbf{c} in projection along $|\textbf{\textrm{r}}|$ when viewed along the directions indicated by the black arrows. The red surfaces represent values of $\chi$ between 0 and 1 and blue surfaces represent values between 0 and $-$1. The numbers on the surfaces in \textbf{c} and \textbf{d} correspond to the molecular arrangements in Figure 8.}
\end{figure}

For phase II, shown in Figures 9\textbf{a} and 9\textbf{b}, only a red isosurface exists since all the molecules are aligned with their normals approximately parallel to the \textit{c}-axis and so $\chi$ ranges between 0 and 1. Each molecular pair is only able to possess a geometrical relationship tending towards scenarios 1 or 2 in Figure 8, and not 3 or 4. The isosurface is restricted to well-defined regions of the 3-D histogram, as would be expected for an ordered structure. At 13\,K, the surface at shortest $|\textbf{r}|$ lies between 4.8 and 5.6\,{\AA} (maximum at 5.2\,{\AA}) and is split over two separate regions with respect to $|\textbf{S}{\cdot}\textbf{r}|$ and $|\textbf{S}^{\prime}{\cdot}\textbf{r}|$; if $|\textbf{S}{\cdot}\textbf{r}|$ or $|\textbf{S}^{\prime}{\cdot}\textbf{r}|$ tends to 1 then the other tends toward 0.65. These regions correspond to four neighbouring molecules --- two above and two below a reference molecule with respect to the \textit{ab} plane, shown in Figure 10\textbf{a}. 

Moving to higher values of $|\textbf{r}|$, the next region of the isosurface then falls between 5.5 -- 6.2\,{\AA} (maximum at 5.9\,{\AA}), which can be attributed to the other two molecules above and below the the central reference molecule, displayed in Figure 10\textbf{b}. Figures 9\textbf{c} and 9\textbf{d} show the remaining nearest neighbours that surround the reference molecule in the \textit{ab} plane --- two at a slightly shorter distance than the other four, which are represented in Figure 9\textbf{a} by the isosurface region 6.0 $> |\textbf{r}| >$ 7\,{\AA}. When the temperature is raised to 176\,K (Figure 9\textbf{b}), the surface is still restricted to the same reasonably well-defined regions but the volume it encloses increases, reflecting the development of thermal disorder in the structure but the retention of the same basic structure and correlations.

\begin{figure}
\begin{indented}
\item \includegraphics[width=8cm]{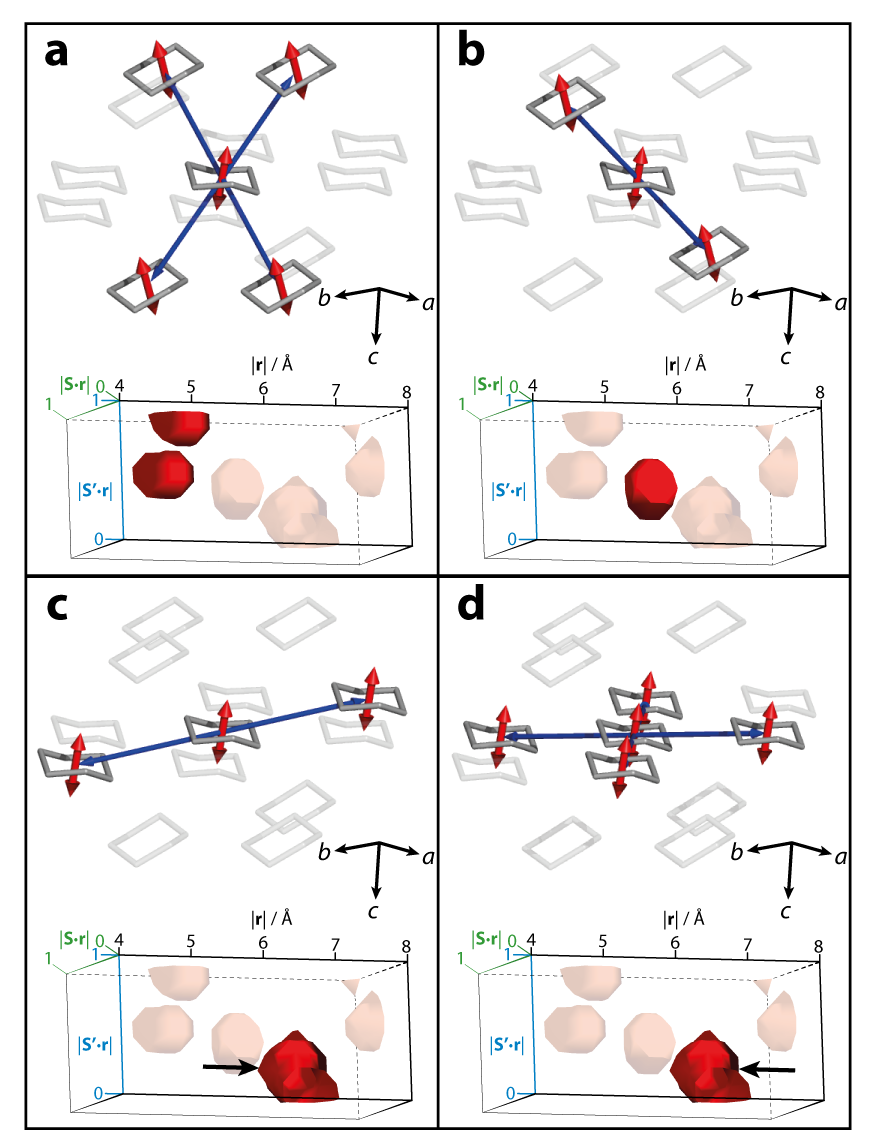}
\end{indented}
\caption{\label{Figure_10.png} Nearest neighbours to a central reference molecule in phase II. Panels \textbf{a} to \textbf{d} show neighbouring molecules at increasing values of $|\textbf{\textrm{r}}|$ respectively. The corresponding regions of the isosurface from Figure 9\textbf{a} are shown at the bottom of each panel.}
\end{figure}
 
In phase I, (Figures 9\textbf{c} and 9\textbf{d}) the molecules exhibit a far wider range of orientations, and so there is a blue isosurface, representing $\chi$ values between $-$1 and 0, in addition to the red. At each extreme end of the blue surface, if $|\textbf{S}{\cdot}\textbf{r}|$ = 1 then $|\textbf{S}^\prime{\cdot}\textbf{r}|$ = 0, and vice versa. The corresponding orientations are depicted in Figure 8 by the equivalent scenarios, 3 and 4. The edges of the blue isosurface have a low $|\textbf{r}|$ value of 5.2\,{\AA}, whereas at intermediate values of $|\textbf{S}{\cdot}\textbf{r}|$ and $|\textbf{S}^\prime{\cdot}\textbf{r}|$ (\textit{ca.}~0.3 -- 0.7), the isosurface does not fall below 5.7\,{\AA} except for a small protrusion in the centre that reaches 5.4\,{\AA}. This shows that when converting between scenarios 3 and 4 the molecules must move further apart in order to spatially accommodate the rotation. That this `conversion' section covers a relatively small region of the 3-D histogram suggests that fewer molecules prefer to adopt the orientation where both molecules make an angle of \textit{ca}.~45$^\circ$ with the centroid-centroid vector. This may represent something akin to an orientational transition state between the presumably sterically favoured edge-to-face molecular arrangements, typified by scenarios 3 and 4. The overall $|\textbf{r}|$ range that the blue isosurface covers (\textit{ca}.~5.2 -- 7.3\,{\AA}) corresponds to pairwise correlations between molecules in the corner of the unit cell and at the face centre.
 
The red isosurface in Figures 9\textbf{c} and 9\textbf{d} is less evenly distributed, where only a small proportion of molecules align with their neighbour and also with their centroid-centroid vector --- numbered `1' in the Figure. Significantly more molecules prefer to be anywhere between being perpendicular to the centroid-centroid vector, numbered `2', or making an angle of \textit{ca.}~45\,$^\circ$ with it. The smaller region of the red surface reaches low-$|\textbf{r}|$ values of 4.8\,{\AA}, which is significantly shorter than the corner-to-face distance seen in the average structure (\textit{cf}.~6.1\,{\AA}), demonstrating that in order for two molecules to move this close together they must adopt a geometric arrangement resembling scenario 1 in Figure 8. This distorted geometry bears similarity to the molecular pair arrangement in phase II with the shortest separation distance (5.241\,{\AA} at 176\,K), represented by Figure 10\textbf{a}.

The larger of the red regions in Figures 9\textbf{c} and 9\textbf{d} lies between $|\textbf{r}| =$ 5.8 and 7.4\,{\AA} --- a much shorter distance than the cubic cell dimension of 8.6836\,{\AA}. This region is most easily considered in two parts --- the first at shorter $|\textbf{r}|$ corresponding to molecules that are approximately located on their average positions in the corner and face centre of the unit cell. The second part, at longer $|\textbf{r}|$ can be attributed to molecules that have deviated from their average positions where the face centre molecules move further away from the cell origin. When this far apart, the molecules experience reduced steric hindrance from each other and so are able to adopt a geometry more closely resembling scenario 2 in Figure 8. Similarly to the smaller red region in Figures 9\textbf{c} and 9\textbf{d}, the distortion from the average structure results in local environments that are reminiscent of the average molecular packing seen in phase II, where the molecules are 6.451\,{\AA} (Figure 10\textbf{c}) and 6.516\,{\AA} (Figure 10\textbf{d}) apart within the \textit{ab} plane.  

A visual comparison of the red isosurfaces at 176 and 266\,K shows that they enclose a broadly similar region of the 3-D histogram (Figure 11). This is consistent with the observations, discussed above, that instantaneous structural distortions in phase I result in local geometry that is more like the average structure of phase II than I. Although it is harder to identify local features in the phase II models that suggest transformation into phase I is developing within the structure (other than increased thermal motion), it is much easier to see how phase I might transform into II. The main difference between the regions of red isosurface in the two phases is their relative sizes, however the area of phase I is necessarily reduced owing to a significant quantity of molecules being able to adopt the perpendicular geometries as a result of the rotational disorder.  

\begin{figure}
\begin{indented}
\item \includegraphics[width=\textwidth]{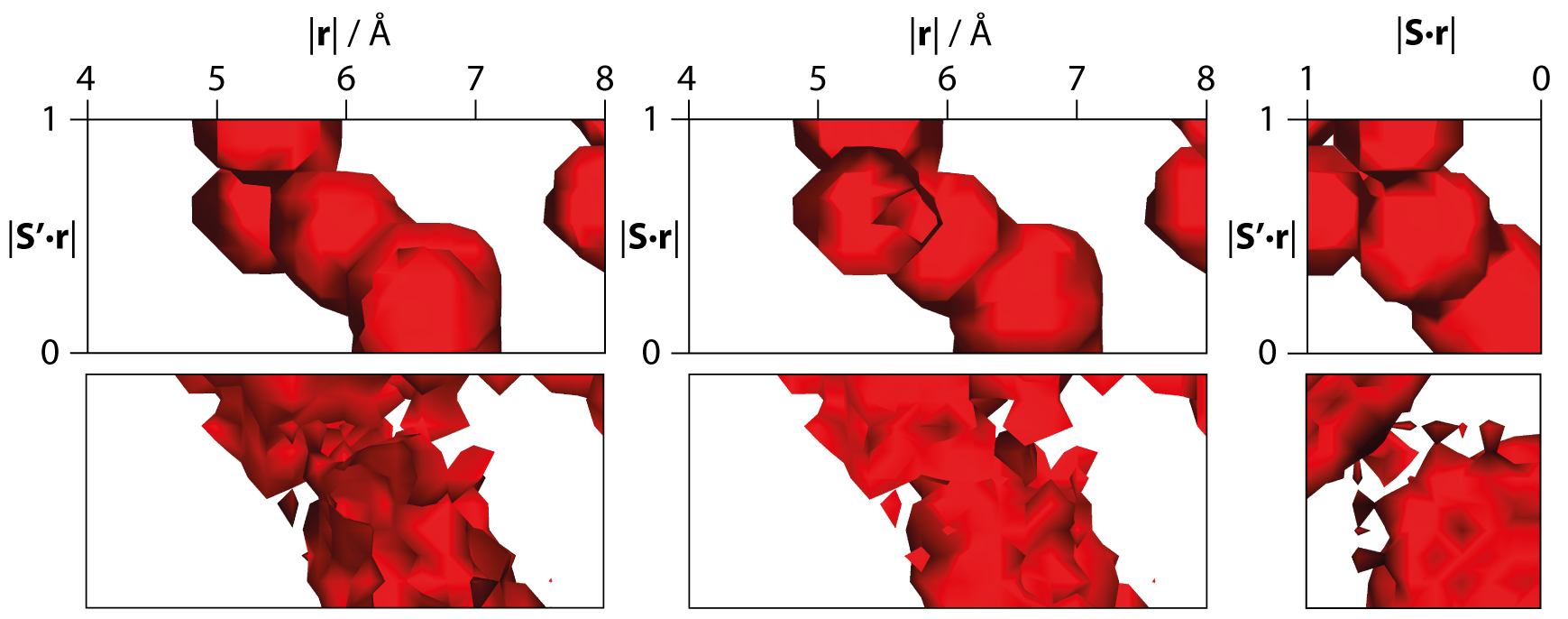}
\end{indented}
\caption{\label{Figure_11.png} Red isosurface at 176\,K (top row) viewed along three different directions. The corresponding orientations of the isosurface at 266\,K are shown along the bottom row. The blue isosurface at 266\,K has been removed for clarity.}
\end{figure}

\subsection{Molecular orientation distribution}
The unit cell vectors of phases I and II are related by the transformation

\ 
\begin{equation}
\left[ \begin{array}{c}
\textbf{a} \\ \textbf{b} \\ \textbf{c} 
\end{array} \right]_{\mathrm{II}} = 
\left[\begin{array}{rrr}
-0.5 & 0.5 & -1 \\
-0.5 & -0.5 & 0 \\
-0.5 & 0.5 & 1 \\
\end{array} \right]
\left[\begin{array}{c}
\textbf{a} \\ \textbf{b} \\ \textbf{c}
\end{array}\right]_{\mathrm{I}}
\end{equation}
\ 

\noindent (see Figure 5 in \cite{Kahn_1973}). The transition involves alignment of each molecular three-fold axis with the monoclinic \textit{c}-axis (parallel to the [$\bar{1}$12] direction in the cubic phase). The distance between molecular centroids changes such that the density in the (001) plane decreases but these layers approach one another more closely in the [001] direction. We have examined the distribution of molecular orientations in both phases, plotted in Figure 12, where the monoclinic cell axes have been transformed to the cubic axis system and had \textit{m}3\textit{m} symmetry applied to the orientational distribution to aid comparison with phase I. Additionally shown in Figure 12 are the orientational distributions for the results of the MD simulations, i.e. the starting atomic coordinates for RMC refinement. The imposition of cubic symmetry generates symmetry equivalent peaks for both data and noise and so it is important not to place too much significance on fine details in the plots.  

In phase II the molecular orientations align preferentially along the $\langle112\rangle$ directions, with respect to the cubic axes, at 13\,K and as temperature is raised to 176\,K, although the distribution of orientations increases, the preferred orientations are still along $\langle112\rangle$. This set of directions corresponds to the \textit{c}-axis in the monoclinic cell setting, showing that the molecules are in a favourable orientation to transform. The plots for phase I are shown on a different scale to phase II as otherwise they appear to show a random distribution of molecular orientations, potentially masking any subtle structural features. However, rescaling the data also enhances the level of noise and so we tentatively interpret the plots as showing that the molecular orientations are not spherically distributed, but instead preferentially align along the $\langle110\rangle$ directions. However further work needs to be carried out to unambiguously establish whether the preferential alignment with the $\langle110\rangle$ directions in phase I is a real feature or not.

\subsection{Molecular dynamics-informed refinements}
Of particular note in Figure 12 is that across the whole temperature range there is very little difference between the distribution of molecular orientations in the MD and RMC models. In phase II, there is not a strong need for the RMC refinement to alter the molecular orientations significantly from those defined by the MD simulations. They are evidently in good agreement with the data to begin with as the refined orientations show little change, yet the fits to the $D(r)$, $F(Q)$ and Bragg profile data (Figure 4) are excellent.

\begin{figure}
\begin{indented}
\item \includegraphics[width=10.8cm]{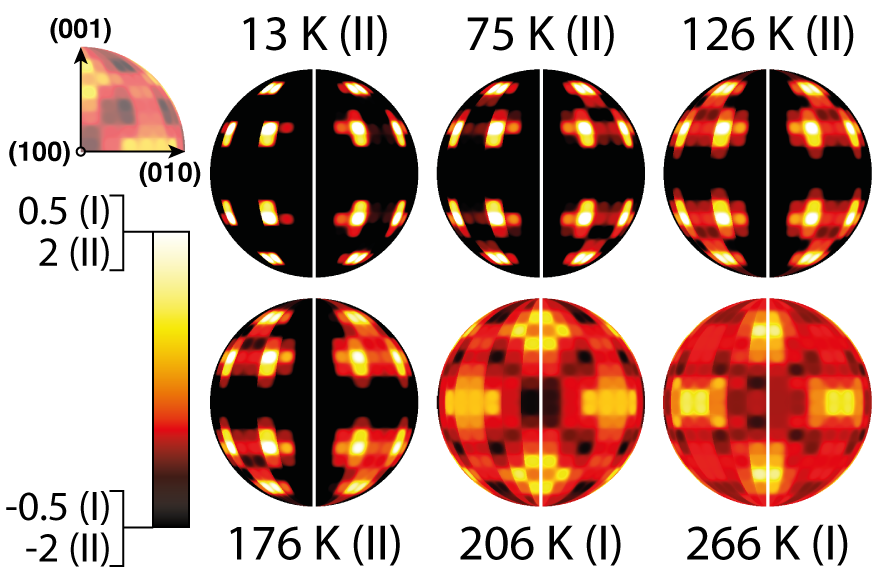}
\end{indented}
\caption{\label{Figure_12.png} Molecular orientation distributions plotted against cubic axes, the directions of which are shown in the upper-left plot. The left and right halves of each distribution correspond to the post-MD, and post-RMC orientations respectively. Light colours represent the directions perpendicular to which the mean cyclohexane planes are oriented. The bar on the lower-left shows the scale on which the data are plotted and the phase is given in parentheses.}
\end{figure}

In phase I, although the overall agreement between the RMC-refined models and the data is still very good, a couple of peak intensities in the Bragg profile are over- and under-fitted. The suggestion was put forward earlier in this manuscript that the disordered cyclohexane models might be challenging for RMC refinement as all molecules would need to be moved cooperatively. Indeed, this can be demonstrated more clearly when starting from an initially ordered model where the molecules have been placed arbitrarily on the face centres of the unit cell but are all oriented in the same direction. Figure 13 shows the RMC fits and the atomic site and molecular orientation distributions for such a model, having been refined for the same period of time as the other phase I structures reported here. The atomic distribution shows that the final model does not meet the FCC symmetry requirements and is accompanied by poor fits to the data. In particular, the additional erroneous peaks in the Bragg profile fit show a lowering of the average symmetry. A representative overall $\chi^2$ value (1) taken from one of the 10 refined models is 25.10 --- over twice the value obtained when starting from the MD coordinates (12.16), suggesting that the model is either jammed in a local minimum or the refinement is very slow to converge. It may be the case that a satisfactory model could be obtained if the refinement were allowed to continue for a greater period of time but this becomes impractical when a reasonable structure can be reached much faster when the MD simulations are utilised. 

\begin{figure}
\begin{indented}
\item \includegraphics[width=10.8cm]{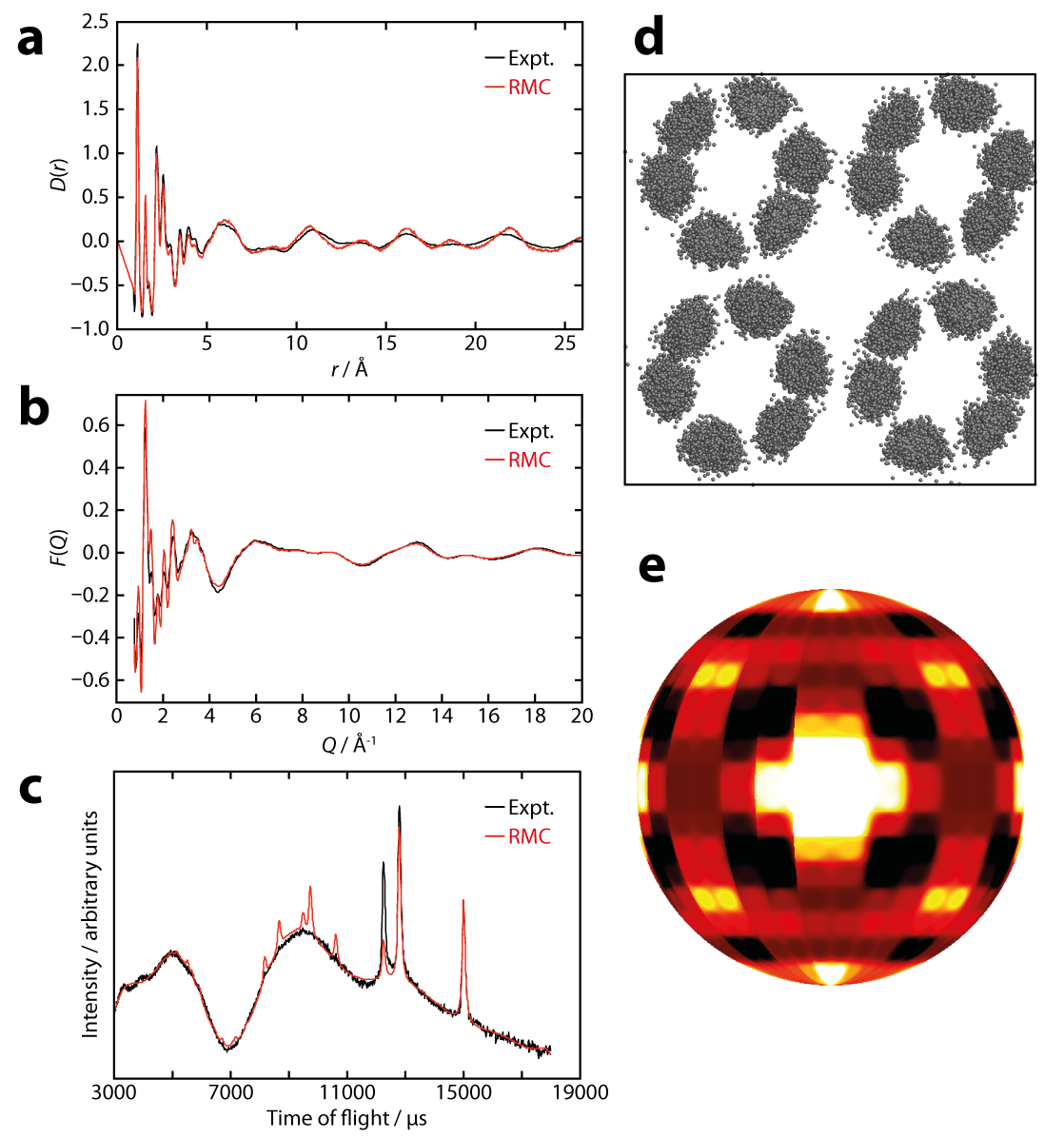}
\end{indented}
\caption{\label{Figure_13.png} RMC fits to the $D(r)$ (\textbf{a}), $F(Q)$ (\textbf{b}) and Bragg profile data (\textbf{c}) at 266\,K when using an ordered starting model. Panels \textbf{d} and \textbf{e} show the atomic site and molecular orientation distributions, respectively.}
\end{figure}

This highlights the usefulness of combining RMC with other techniques --- MD in this case --- for arriving at a structure that is consistent with the experimental total scattering data. Other examples in the literature have also demonstrated that RMC refinements for `problematic systems' arrive at more sensible structures when aided by complementary information from other techniques \cite{Cliffe_2010,Paddison_2012}. Cyclohexane is arguably another challenging example as the correlated motion has to be dealt with as well as preservation of the relatively complex molecular geometry (in the context of other RMC-refined systems). Here, the role of the MD is to deal with the orientations of the molecules, and the RMC to broaden this distribution. The more significant outcome of the RMC refinement is that the intermolecular distances are successfully refined to show excellent agreement with the experimental data, whereas the separation distances predicted by the MD are relatively poor (Figure 3), especially in phase I.

\section{Conclusion}
Neutron total scattering data have been collected to a much higher \textit{Q}$_\mathrm{max}$ (45\,{\AA}$^{-1}$) than previously reported experiments, reducing the effects of Fourier truncation. The crystalline and plastic phases of cyclohexane have been modelled using RMCProfile and MD simulations, forming one of the most challenging systems to have been attempted with the RMC method in terms of the molecular geometry and complexity of intermolecular steric interactions. From the resulting atomic configurations, 3-D correlation functions have been extracted for both phases that confirm molecular orientation is coupled to spatial distribution. Similar local packing is observed in both phases, and features from the average structure of phase II are seen to exist locally in instantaneously distorted regions of phase I.

Plots of molecular orientation distributions have shown that the molecular arrangements are largely defined by the MD simulations and that the distributions of orientations are then broadened by the RMC refinements. The main contribution of the RMC process to the resulting atomic configurations is to refine the molecular separation distances such that they show excellent agreement with the experimental data across all the investigated temperature points.

\section{Acknowledgements}
We thank the EPSRC for funding to NPF and SP and to ALG (grant code: EP/G004528/2), the ERC for funding to NPF and ALG (grant code: 279705), and the STFC for provision of neutron beamtime.


\section{References}
\begin{thebibliography}{46}

\bibitem{Sun_2008}
Sun Z, Zhou J, Blomqvist A, Xu, L and Ahuja R 2008 {\it J. Phys. Condens. Matter} {\bf 20} 205102/1

\bibitem{Caravati_2010}
Caravati S, Bernasconi M and Parrinello M 2010 {\it J. Phys. Condens. Matter} {\bf 22} 315801/1

\bibitem{Chemburkar_2000}
Chemburkar S R \textit{et al}, 2000 {\it Org. Process Res. Dev.} {\bf 4} 413

\bibitem{Wood_2006}
Wood P A, Forgan R S, Henderson D, Parsons S, Pidcock E, Tasker P A and Warren J E 2006 {\it Acta Crystallogr., Sect. B Struct. Sci.} {\bf B62} 1099

\bibitem{Rivera_2008}
Rivera S A, Allis D G and Hudson B S 2008 {\it Cryst. Growth Des.} {\bf 8} 3905

\bibitem{Funnell_2011}
Funnell N P, Marshall W G and Parsons S 2011 {\it CrystEngComm} {\bf 13} 5841

\bibitem{Johnstone_2008}
Johnstone R D L, Francis D, Lennie A R, Marshall W G, Moggach S A, Parsons S, Pidcock E and Warren J E 2008 {\it CrystEngComm} {\bf 10} 1758

\bibitem{Hui_2005}
Hui Q, Tucker M G, Dove M T, Wells S A and Keen D A 2005 {\it J. Phys. Condens. Matter} {\textbf 17} S111

\bibitem{Tucker_2000}
Tucker M G, Dove M T and Keen D A 2000 {\it J. Phys. Condens. Matter} {\bf 12} L723

\bibitem{Tucker_2001}
Tucker M G, Squires M P, Dove M T and Keen D A 2001 {\it J. Phys. Condens. Matter} {\bf 13} 403

\bibitem{Shoemaker_2011}
Shoemaker D P, Llobet A, Tachibana M and Seshadri R 2011 {\it J. Phys. Condens. Matter} {\bf 23} 315404/1

\bibitem{Dove_2002}
Dove M T, Tucker M G and Keen D A 2002, {\it Eur. J. Mineral} {\bf 14} 331

\bibitem{Temleitner_2010}
Temleitner L and Pusztai L 2010 {\it Phys. Rev. B Condens. Matter Mater. Phys.} {\bf 81} 134101/1

\bibitem{Billinge_2010}
Billinge S J L, Dykhne T, Juh\'{a}s P, Bo\u{z}in E, Taylor R, Florence A J and Shankland K 2010 {\it CrystEngComm} {\bf 12} 1366

\bibitem{Dykhne_2011}
Dykhne T, Taylor R, Florence A and Billinge S J L 2011 {\it Pharm. Res.} {\bf 28} 1041

\bibitem{Rademacher_2012}
Rademacher N, Daemen L L, Chronister E L and Proffen T 2012 {\it J. Appl. Crystallogr.} {\bf 45} 482  

\bibitem{Farman_1987}
Farman H, Dore J C, Bellissent-Funel M C and Montague D G {\it Mol. Phys.} {\bf 61} 583

\bibitem{Farman_1991}
Farman H, O'Mard L, Dore J C and Bellissent-Funel M C 1991 {\it Mol. Phys.} {\bf 73} 855

\bibitem{Farman_1996}
Farman H, Coveney F, Dore J C and Bellissent-Funel M C 1996 {\it Mol. Phys.} {\bf 87} 1217

\bibitem{Kahn_1973}
Kahn R, Fourme R, Andre D and Renaud M 1973 {\it Acta Crystallogr., Sect. B} {\bf 29} 131

\bibitem{Bansal_1979}
Bansal M L and Roy A P 1979 {\it Mol. Phys.} {\bf 38} 1419

\bibitem{Brodka_1992}
Brodka A and Zerda T W 1992 {\it J. Chem. Phys.} {\bf 97} 5669

\bibitem{Ehrenberg_1992}
Ehrenberg V, Hoheisel C and W{\"u}rflinger A 1992 {\it Phys. Chem. Liq.} {\bf 25} 27

\bibitem{McGrath_1997}
McGrath K J and Weiss R G 1997 {\it Langmuir} {\bf 13} 4474

\bibitem{McGreevy_1988}
McGreevy R L and Pusztai L 1988 {\it Mol. Simul.} {\bf 1} 359

\bibitem{Ibberson_1996}
Ibberson R M 1996 {\it J. Appl. Crystallogr.} {\bf 29} 498

\bibitem{Williams_1998}
Williams W G, Ibberson R M, Day P and Enderby J E 1998 {\it Physical B (Amsterdam)} 241-243 234

\bibitem{McLain_GUDRUN}
McLain S E, Bowron D T, Hannon A C and Soper A K {\it GUDRUN, a Computer Program Developed for Analysis of Neutron Diffraction Data}, ISIS Facility, Rutherford Appleton Laboratory, Chilton, UK

\bibitem{Keen_2001}
Keen D A 2001 {\it J. Appl. Crystallogr.} {\bf 34} 172

\bibitem{Larson_GSAS}
Larson A C and von Dreele R B 2000 {\it General Structure Analysis System (GSAS)}, Los Alamos National Laboratory

\bibitem{Toby_2001}
Toby B H 2001 {\it J. Appl. Crystallogr.} {\bf 34} 210

\bibitem{Bruno_2002}
Bruno I J, Cole J C, Edgington P R, Kessler M, Macrae C F, McCabe P, Pearson J and Taylor R 2002 {\it Acta Crystallogr., Sect. B Struct. Sci.} {\bf B58} 389

\bibitem{Dauber_1988}
Dauber-Osguthorpe P, Roberts V A, Osguthorpe D J, Wolff J, Genest M and Hagler A T 1988 {\it Proteins Struct. Funct. Genet.} {\bf 4} 31

\bibitem{Berendsen_1984}
Berendsen H J C, Postma J P M, Van Gunsteren W F, DiNola A and Haak J R 1984 {\it J. Chem. Phys.} {\bf 81} 3684

\bibitem{Parrinello_1982}
Parrinello M and Rahman A, 1982 {\it J. Chem. Phys.} {\bf 76} 2662

\bibitem{Delley_1990}
Delley B 1990 {\it J. Chem. Phys.} {\bf 92} 508

\bibitem{Tucker_2007}
Tucker M G, Keen D A, Dove M T, Goodwin A L and Hui Q 2007 {\it J. Phys. Condens. Matter} {\bf 19} 335218/1

\bibitem{Crystalmaker}
{\it CrystalMaker{\textsuperscript \textregistered}, CrystalMaker Software Limited}, http://www.crystalmaker.com

\bibitem{Macrae_2008}
Macrae C F, Bruno I J, Chisholm J A, Edgington P R, McCabe P, Pidcock E, Rodriguez-Monge L, Taylor R, van der Streek J and Wood P A 2008 {\it J. Appl. Crystallogr.} {\bf 41} 466

\bibitem{Li_2003}
Li J 2003 {\it Modelling Simul. Mater. Sci. Eng.} {\bf 11} 173

\bibitem{Pymol}
{\it The Pymol Molecular Graphics System, Version 1.5.0.4}, Schr{\"o}dinger, LLC

\bibitem{Henderson_2007}
Henderson A 2007 {\it Paraview Guide, A Parallel Visualization Application} Kitware Inc. 

\bibitem{Hampton_1976}
Hampton E M and Sherwood J N 1976 {\it J. Chem. Soc., Faraday Trans. 1} {\bf 72} 2398

\bibitem{Wood_2008}
Wood P A, McKinnon J J, Parsons S, Pidcock E and Spackman M A 2008 {\it CrystEngComm} {\bf 10} 368

\bibitem{Cliffe_2010}
Cliffe M J, Dove M T, Drabold D A and Goodwin A L 2010 {\it Phys. Rev. Lett.} {\bf 104} 125501

\bibitem{Paddison_2012}
Paddison J A M and Goodwin A L 2012 {\it Phys. Rev. Lett.} {\bf 108} 017204

\end{thebibliography}
\end{document}